\documentclass{ws-procs10x7}

\begin{document}

\title{Jets in Photoproduction and in the Transition Region to DIS at the HERA collider}

\author{Juraj Bracinik}

\address{Max-Planck-Institute for Physics (Werner Heisenberg Institute),
F\"ohringer Ring 6 \\ 80805 M\"unchen, Germany
\\E-mail: bracinik@mppmu.mpg.de \\ For H1 and ZEUS collaborations}

\twocolumn[\maketitle\abstract{
Recent results on jets in photoproduction and in deep-inelastic scattering at
low $Q^2$ by the H1 and
ZEUS collaborations are reviewed.
}]

\section{Introduction}
 
The photon is probably the best known elementary particle. It is a quantum 
of the gauge field, and as such it is considered to be massless, charge-less and to couple
point-like.

While interactions of photons with leptons are described by QED with high
precision, interactions of photons with hadrons still bring surprises. This is
caused by the fact, that in the same way, as the photon can fluctuate into
an electron-positron pair, it can fluctuate into a pair of quark and anti-quark,
which interact strongly. The photon then behaves as a hadron.

In Leading Order (LO), it is possible to distinguish between direct processes
(the photon interacts electromagnetically) and resolved processes (the photon fluctuates
into a partonic system, of which one of the partons then interacts). Beyond LO this
classification becomes ambiguous.

\section{Theoretical description of photon-proton interactions}

Perturbative QCD calculations which aim 
to describe interactions of photons with protons use
as an input parton density functions of the proton (obtained from global fits to
DIS and hadron collision data) and of the photon (extracted from data on
$\gamma \gamma$ collisions).

Depending on the way perturbative QCD is used, there are two groups of
approaches.

Leading order plus Parton Shower (LO+PS) models 
 combine LO matrix
elements with parton showers re-summing the leading logarithmic contributions from all 
orders.\cite{h1_dijets_herwig}
 Hadronization is included using a QCD motivated phenomenological model and predictions are directly
compared to data on hadron level.

Next-to-leading order (NLO) calculations use matrix elements up to a fixed order in
$\alpha_S$ (for most processes up to 
$\alpha_S^2$).\cite{h1_incl_frixione_ridolfi,zeus_incl_klasen,zeus_dijets_graudenz}
 They provide
predictions on parton level. Before comparing to data, NLO predictions are
corrected for hadronization effects. These are estimated using LO+PS models discussed above.

\section{Experimental conditions}

Depending on how events are selected, we distinguish between  tagged
photoproduction (scattered electron is measured in downstream calorimeter, $Q^2
\le 10^{-2} \; {\rm GeV}^2$), un-tagged photoproduction (electron is 
not observed in main detector, $Q^2
\le 1 \; {\rm GeV}^2$) and low $Q^2$ region (electron is measured in main detector,  
$Q^2 \ge 2 \; {\rm GeV}^2$). The distribution of the center-of-mass energy of the $\gamma p$ system
depends on the exact event selection; at HERA it extends up to $280 \; {\rm GeV}$. 

The results are presented in the hadronic CMS (center-of-mass frame of
$\gamma p$ system), for jet finding, an inclusive $k_{t}$ algorithm is
used\cite{ktclus} in the same frame.

Before being compared to theory, data are corrected to hadron level, i.e. acceptance and 
effects due to the detector and the reconstruction software are corrected for. These
corrections are calculated using LO+PS models together with a detailed detector
simulation.

\section{Inclusive jets}
\begin{figure}
\epsfxsize140pt
\figurebox{}{}{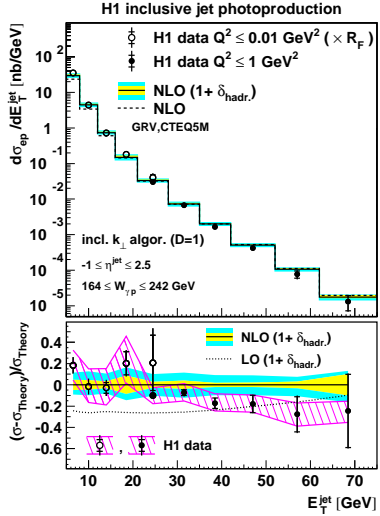}
\caption{Cross section of inclusive jets in photoproduction as a function of 
$E_T^{jet}$ from H1.}
\label{fig:incl_et}
\end{figure}

To check our understanding of jet production in photoproduction one can simply
count the number of jets as a function of their transverse energy. The cross section of
inclusive jets in the pseudo-rapidity range $-1 < \eta < 2.5$
has been measured by both H1
(Fig.~\ref{fig:incl_et}) and ZEUS.\cite{h1_inclusive,zeus_inclusive}

One can see an excellent agreement between the NLO calculation and the data.
Agreement extends down to low value of $E_T$ ($5 \; {\rm GeV}$), where
hadronization corrections (including effects of the underlying event) become
significant. The dominant experimental error is coming from the energy scale
uncertainty, which is smaller then the renormalization scale uncertainty of the theory. 

\section{Dijets in photoproduction and at low $Q^2$ DIS}
\begin{figure}
\begin{center}
\epsfxsize96pt
\epsfysize150pt
\figurebox{}{}{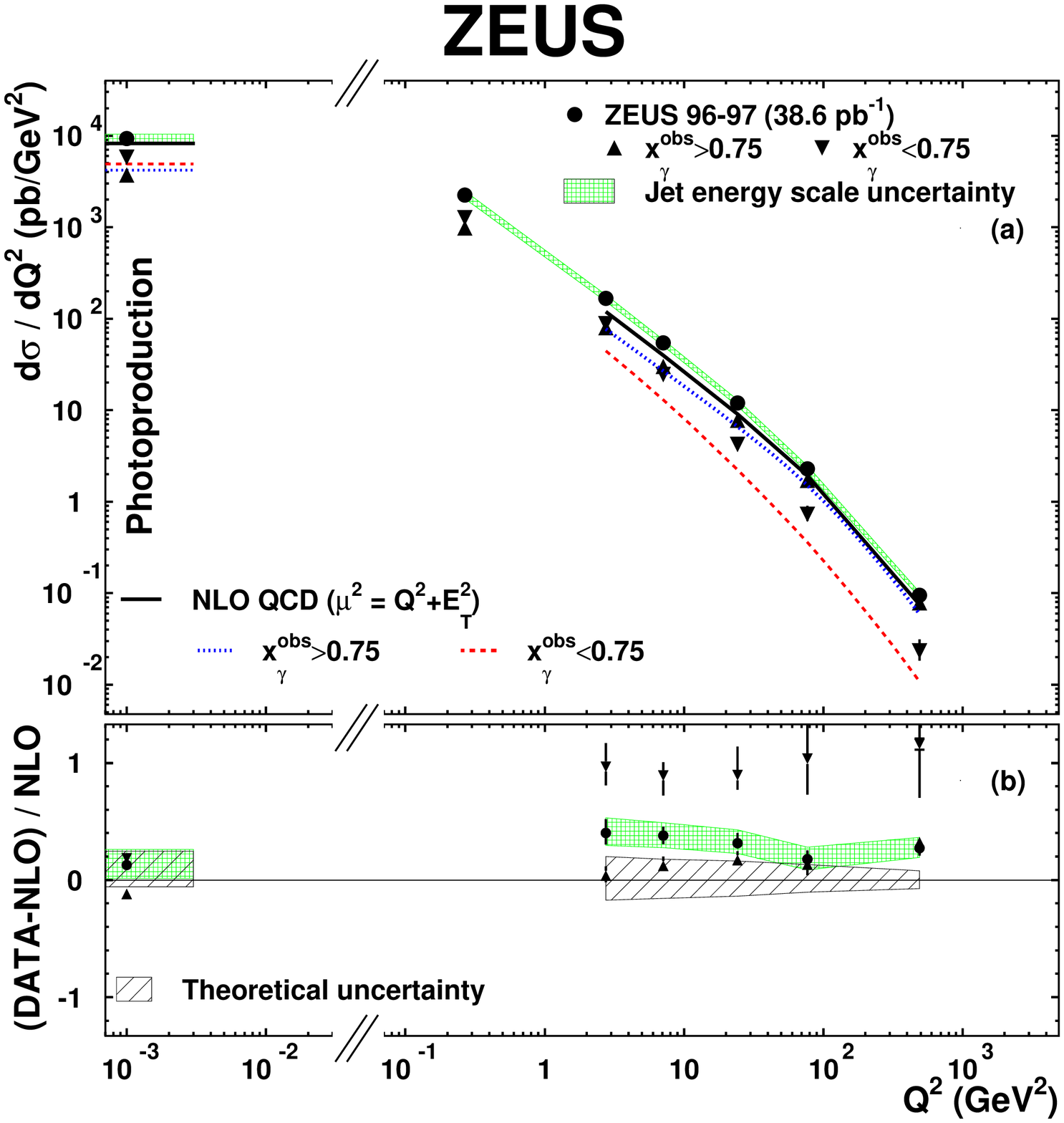}
\epsfxsize96pt
\epsfysize150pt
\figurebox{}{}{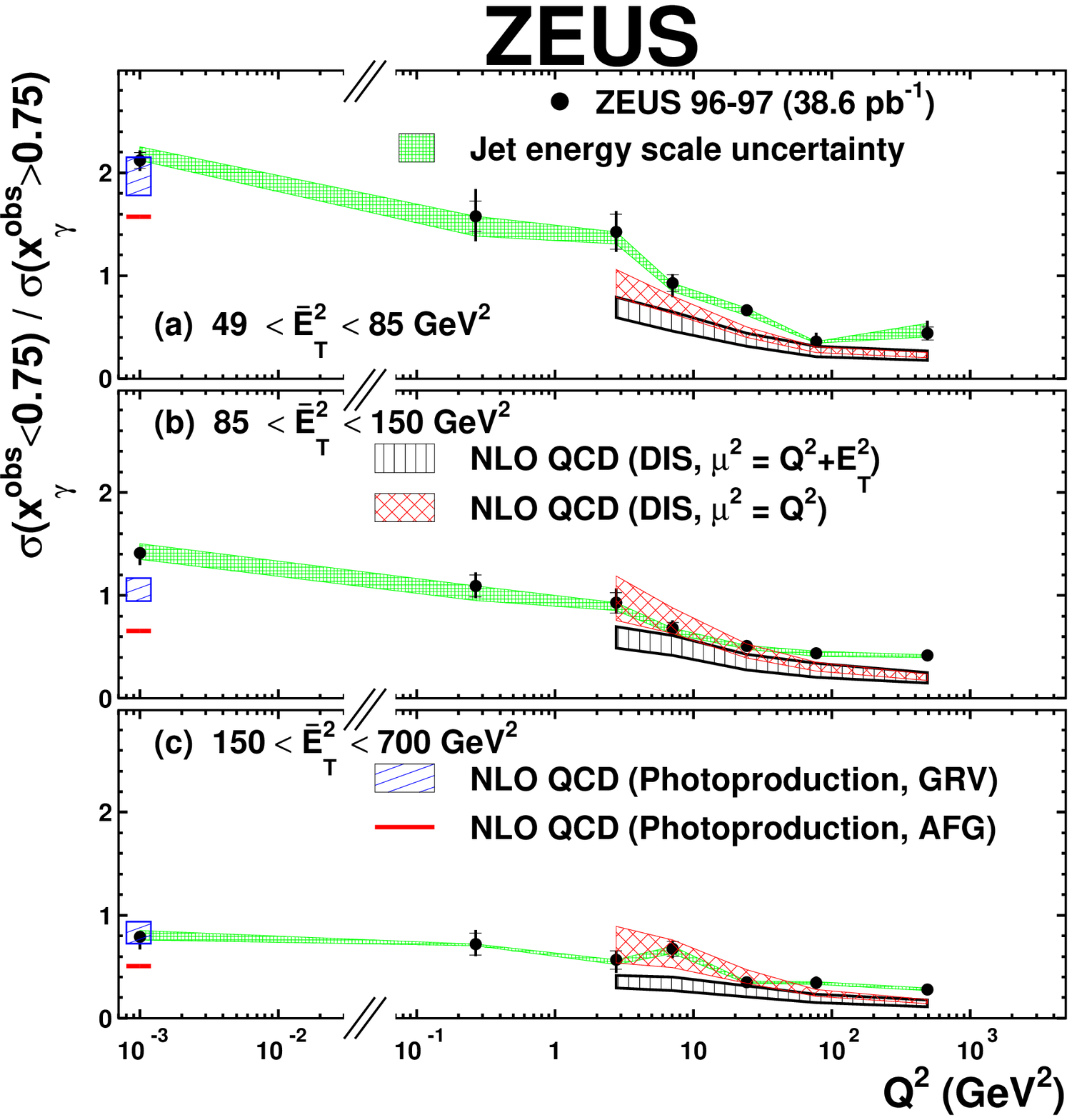}
\end{center}
\caption{Dijet production in photoproduction and DIS from ZEUS
as a function of $Q^2$, cross section (left), ratio between 
cross sections for resolved and direct event sample (right).  }
\label{fig:dijets_zeus_nlo}
\end{figure}
A measurement of inclusive jet production has clear advantages. Theoretical
predictions are "safe", the measurement is least restrictive in phase space and 
offers good statistical precision. On the other hand, dijets allow to construct
more differential quantities, for example 
$x_{\gamma}^{obs}$.\footnote {defined as 
$x_{\gamma}^{obs} = \sum_{jets} (E_j^* - p_z^*) / \sum_{hadrons} (E_j^* - p_z^*)$}
At parton level in LO this variable corresponds to the fractional photon energy
of the parton "in" the photon 
entering the hard subprocess. Higher orders, hadronization and detector 
resolution smear-out the correlation, but still one expects for direct processes
$x_{\gamma}^{obs}$ to be close to one and significantly smaller than one for resolved
processes. 
 
In
photoproduction ($Q^2 = 0$) a photon behaves part of the time as a hadron, 
which is manifested 
by the presence
of its resolved part. On the other hand, in DIS, at high enough $Q^2$, 
the photon is point-like.
What happens in the transition region?

ZEUS has presented the cross section on dijets in photoproduction
and DIS (Fig.~\ref{fig:dijets_zeus_nlo} left)
in the hadronic CMS, in the  pseudo-rapidity range $-3 < \eta^* < 0$ and for
$E_T> 7.5 \; (6.5) \;  {\rm GeV}$.\cite{zeus_dijets}
While in photoproduction the cross section of dijet production is in agreement with 
NLO,\cite{h1_incl_frixione_ridolfi} in DIS however, the NLO assuming only
pointlike photon\cite{zeus_dijets_graudenz}, underestimates the cross section.

Requiring
$x_{\gamma}^{obs} > 0.75 $($x_{\gamma}^{obs} < 0.75 $), it is possible to enhance
direct (resolved) processes. The sample with predominantly direct
processes is well described by  NLO, while a discrepancy is observed at small
$x_{\gamma}^{obs}$.

     The same data are shown in Fig.~\ref{fig:dijets_zeus_nlo} (right side) 
in the form of the ratio 
$R=\sigma(x_{\gamma}^{obs} < 0.75)/\sigma(x_{\gamma}^{obs} > 0.75)$. In this
ratio correlated experimental and theoretical uncertainties partly cancel. We can see
that the discrepancy between data and direct NLO extends up to rather high $Q^2$
values and is most remarkable at low $E_T$. A change of scale (using $Q^2$
instead of $Q^2 + E_T^2$), improves the agreement at low $Q^2$, but not at higher $Q^2$
values.  

This result may be a hint that higher orders are needed in the perturbative calculation. At low
$Q^2$, it is possible to include them effectively using the concept of resolved
virtual photons.
\begin{figure}
\epsfxsize160pt
\figurebox{}{}{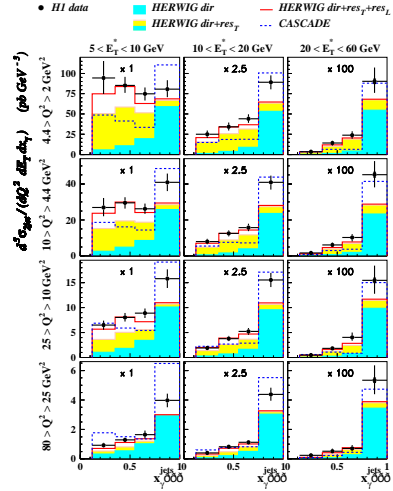}
\caption{Triple differential cross section of dijet production in DIS from H1
compared to HERWIG and CASCADE}
\label{fig:dijets_h1_nlo}
\end{figure}

H1 has measured the triple differential cross section of dijets as
a function of $Q^2$, $E_T$ and $x_{\gamma}$ in DIS at low $Q^2$
in the region $-2.5 < \eta^* < 0$ and $E_T > 7 \; (5) \; {\rm GeV}$.\cite{h1_dijets} 
The comparison of the data with NLO shows that
NLO underestimates the cross section, the discrepancy being most clearly visible
for low $Q^2$, $E_T$ and $x_{\gamma}$. The inclusion of a resolved transverse virtual
photon component in  NLO reduces the discrepancy, but agreement with data is still
not perfect. 

The same data are compared to the LO+PS model of 
HERWIG\cite{h1_dijets_herwig}
in Fig.~\ref{fig:dijets_h1_nlo}. Again, direct
processes alone underestimate the data and inclusion of transverse resolved photons
improves the agreement. Including, in addition a
contribution from resolved longitudinal photons yields an even better description..

An alternative approach to modeling $\gamma p$ interactions is represented by the
CASCADE model,\cite{h1_dijets_cascade2}
which is 
based on the CCFM evolution, with angle ordering instead of
$k_T$ ordering of the radiated gluons and using un-integrated parton densities
of the proton. 
This model is in
reasonable (but not perfect) agreement with the data. This is remarkable, as
CASCADE does not use any concept of a resolved photon; low $x_{\gamma}^{obs}$
events are produced by different evolution from the proton side. 

%
\section{Study of color dynamics in three jet events in photoproduction}
\begin{figure}
\epsfxsize120pt
\figurebox{120pt}{160pt}{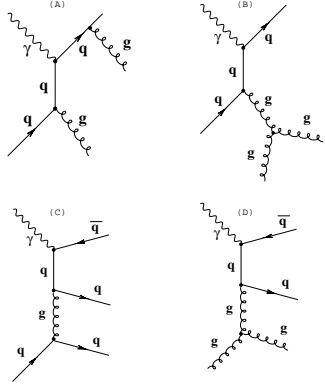}
\caption{LO diagrams for direct three jet events}
\label{fig:threejets_diagrams}
\end{figure}
 
Three jet events are interesting as they feel the triple gluon vertex. It would be 
nice to find three-jet observables sensitive to the structure of the gauge group
behind the strong interaction.

ZEUS has measured cross sections of three-jet production as a
function of angles $\theta_H$, $\alpha_{23}$ and $\beta_{KSW}$.
Jets with  $-1 < \eta < 2.5$ and $E_T > 14 \; {\rm GeV} $ were
selected.\cite{zeus_threejets}
They are 
ordered according to their $E_T$. Then $\theta_H$ is defined as the angle between
the plane defined by the beam axis and jet $1$ and the plane defined by jets $2$ and $3$, 
$\alpha_{23}$  is the angle between jets $2$ and $3$. 

If predominantly direct events are selected ($x_{\gamma} > 0.7$), 
we have in LO
four terms (see Fig.~\ref{fig:threejets_diagrams}). 
Detailed analysis  shows that cross sections
plotted as a function of  $\theta_H$, $\alpha_{23}$ and $\beta_{KSW}$ are
sensitive to the presence of the triple gluon vertex.  The shape of term B (triple gluon
vertex in quark induced events) is different from the other three terms. 
\cite{zeus_threejets}
\begin{figure}
\begin{center}
\epsfxsize96pt
\figurebox{}{}{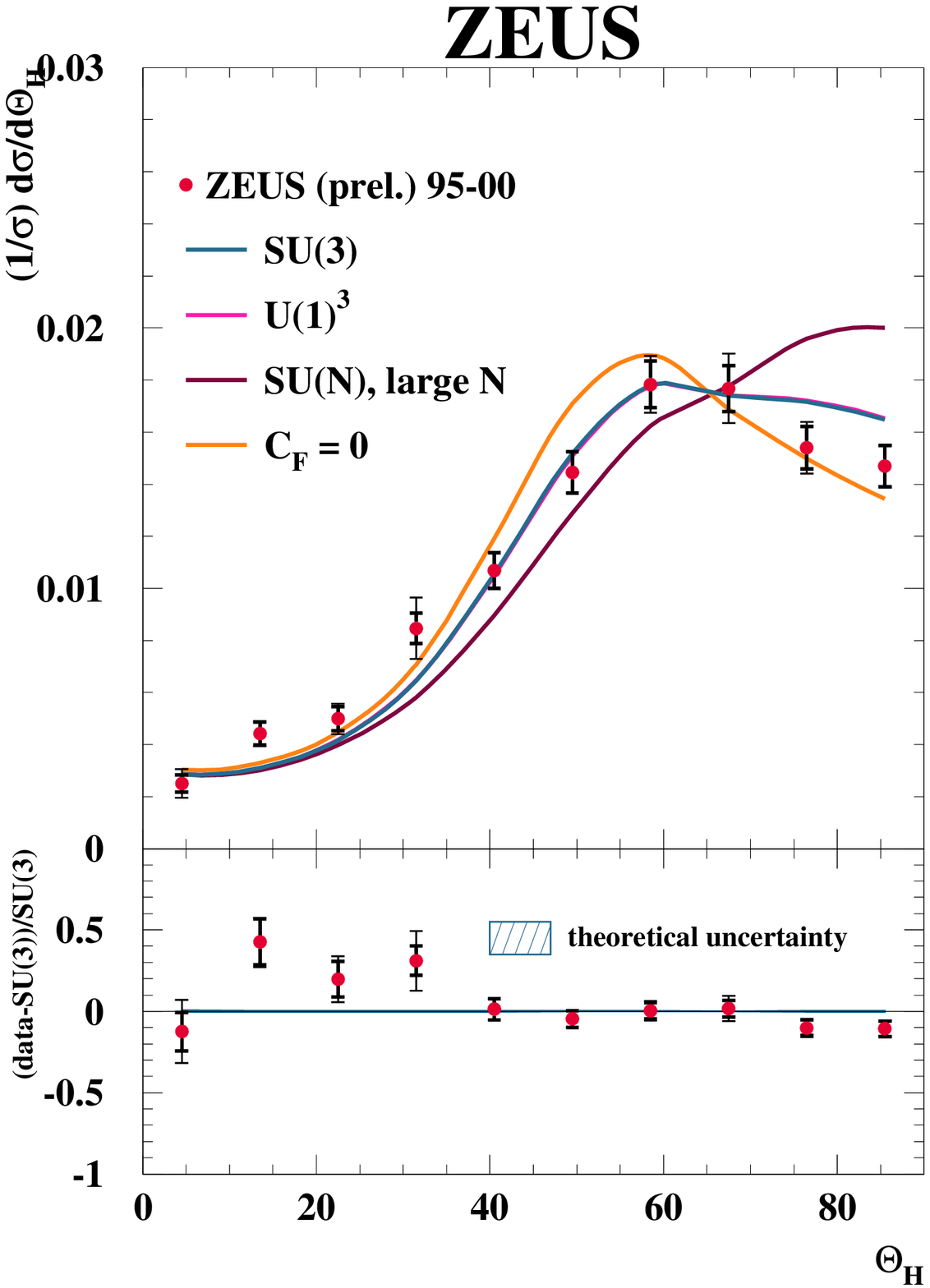}
\epsfxsize96pt
\figurebox{}{}{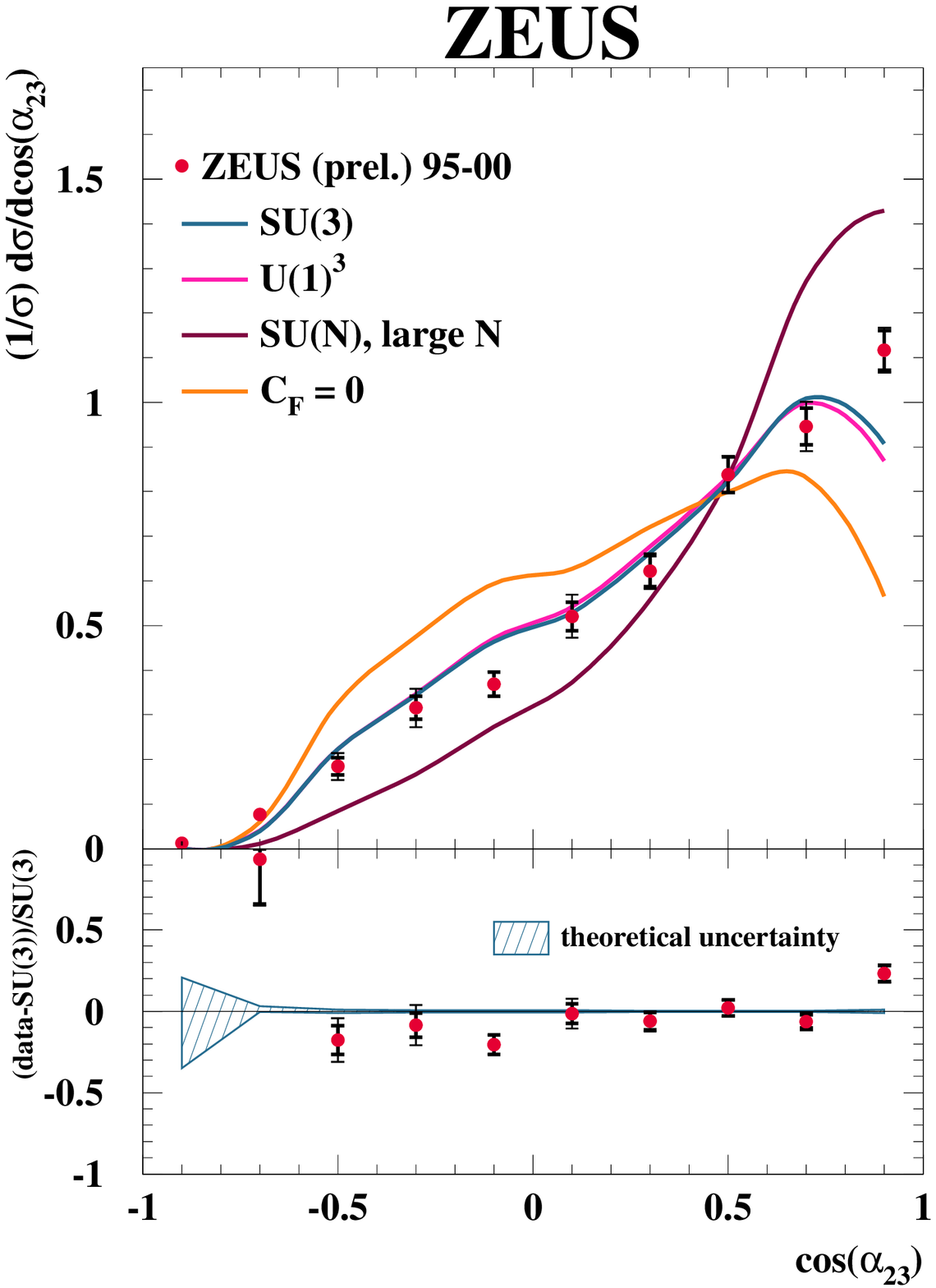}
\end{center}
\caption{Cross section of direct three-jet photoproduction from ZEUS
as a function of $\theta_H$ and  $\cos {\alpha_{23}}$. }
\label{fig:threejets_zeus}
\end{figure}
 
A comparison of data with LO predictions\cite{zeus_incl_klasen} is shown in 
Fig.~\ref{fig:threejets_zeus}. There is
very good agreement with SU(3) expectations. In addition, it is possible to
rule out several exotic possibilities (like $SU(N)$ for large $N$, or $C_F=0$).
Current precision does not allow to distinguish between SU(3) and Abelian case,
because SU(3) predicts only $10 \% 
$ probability for events with a triple gluon vertex in quark induced events.

\section{Summary}
 
Measured cross sections for inclusive jets in photoproduction are in excellent
agreement with NLO. Study of dijets at low $Q^2$ show discrepancies between data
and NLO, indicating the need for higher orders in the perturbative expansion. Inclusion
of resolved longitudinal virtual photon component in low $Q^2$ DIS improves 
the description of the data.

Three-jet events are sensitive to the triple gluon vertex, allowing to study the gauge
structure of the strong interaction. The data are in agreement with LO QCD
predictions. Current precision does not allow to discriminate between SU(3) and the
Abelian case.

\end{document}